\pdfoutput=1

\documentclass[11pt]{article}
\usepackage{amsmath,amsfonts,bm}
\usepackage{microtype}
\usepackage{graphicx}
\usepackage{footmisc}
\usepackage{diagbox}
\usepackage{enumitem}
\usepackage{subfigure}
\usepackage{multirow}
\usepackage{booktabs} 
\usepackage[linesnumbered,ruled]{algorithm2e}
\usepackage{hyperref}
\usepackage{balance}
\usepackage{bm}
\usepackage{enumitem}

\newcommand{\ProName}{ConFRec}
\newcommand{\TAFA}{TAFA} 

\AtBeginDocument{%
  \providecommand\BibTeX{{%
    \normalfont B\kern-0.5em{\scshape i\kern-0.25em b}\kern-0.8em\TeX}}}

\usepackage{emnlp2021}

\usepackage{times}
\usepackage{latexsym}

\usepackage[T1]{fontenc}

\usepackage[utf8]{inputenc}

\usepackage{microtype}

\title{Content Filtering Enriched GNN Framework for News Recommendation}

\author{Yong Gao\textsuperscript{\rm 1}, Huifeng Guo\textsuperscript{\rm 1}\thanks{\ \  Corresponding author}, DanDan Lin\textsuperscript{\rm 2}, Yingxue Zhang\textsuperscript{\rm 3}, Ruiming Tang\textsuperscript{\rm 4} Xiuqiang He\textsuperscript{\rm 5} \\
\textsuperscript{\rm 1, \rm 4, \rm 5}Noah's Ark Research Lab, Huawei, China \\
\textsuperscript{\rm 2}The Hong Kong University of Science and Technology, China \\
\textsuperscript{\rm 3}Noah's Ark Research Lab, Montreal, QC, Canada\\
\{ygaoneeson,lindandan1015\}@gmail.com, \\ \{huifeng.guo, yingxue.zhang, tangruiming, hexiuqiang1\}@huawei.com}

\begin{document}
\maketitle

\begin{abstract}
Learning accurate users and news representations is critical for news recommendation. Despite great progress, existing methods seem to have a strong bias towards content representation or just capture collaborative filtering relationship. 
However, these approaches may suffer from the data sparsity problem (user-news interactive behavior sparsity problem) or maybe affected more by news (or user) with high popularity. 
In this paper, to address such limitations, we propose content filtering enriched GNN framework for news recommendation, \ProName~in short. It is compatible with existing GNN-based approaches for news recommendation and can capture both collaborative and content filtering information simultaneously. 
Comprehensive experiments are conducted to demonstrate the effectiveness of \ProName~over the state-of-the-art baseline models for news recommendation on real-world datasets for news recommendation.
\end{abstract}

\section{Introduction}\label{sec:intro}
Recently, news recommendation platforms have sprung up like mushrooms and are growing rapidly, such as Google News, Bing News and Toutiao.com. Massive news information is produced continuously and it is impractical for users to read all these news due to the time limit. 
Therefore, in online news platforms, it is critical to tackle the information overload and the news recommendation has been playing an increasingly important role  to help users find their interested contents~\cite{liu2010personalized}. 
\begin{figure}[!t]
  \centering
  \includegraphics[width=.46\textwidth]{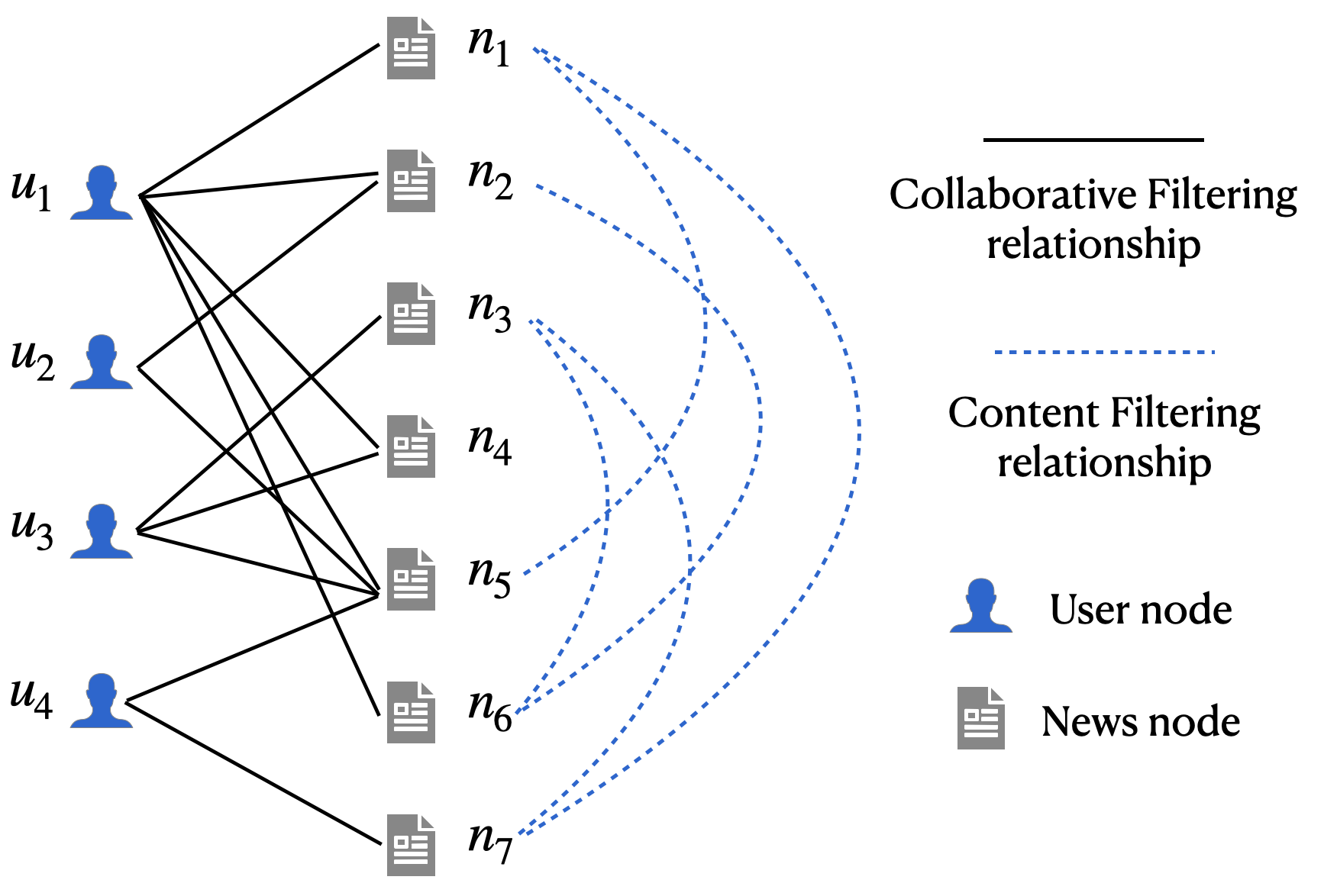}
  \caption{Collaborative filtering and content filtering.}
  \label{CF_CF}
\end{figure}
Almost existing news recommendation methods try to construct personalized news ranking by learning accurate and informative user and news representations (noted as representation-based methods). 
They usually learn news representation from news contents and then construct user representation based on user's history behavior, e.g, aggregating information from the user's clicked news.
For instance, Wang et al. proposed DKN~\cite{wang2018dkn} to learn news representation from news contents via multi-channel CNN and user's representation by aggregating her click news  with different weights. 
Furthermore, DAN~\cite{DAN} and NPA~\cite{NPA} utilized an attention network to identify the important clicked news for generating better user representations. Compared with traditional collaborative filtering methods~\cite{koren2009MF}, which suffer from data sparsity, these methods
have been improved by building semantic news representations directly from news contents. 
However, when the news contents are short and the user historical behavior is sparse, it is difficult to learn reasonable news and user representations~\cite{GERL}. 

In addition to these representation-based methods, several graph neural network (GNN) based approaches, such as GNUD~\cite{GNUD} and GERL~\cite{GERL}, leveraged user-news interactions to capture high-order relatedness between users and news. However, if we only consider the collaborative filtering relationship, the data sparsity problem still exists.
As shown in Figure~\ref{CF_CF}, $u_4$ has two 1-hop neighbors $n_5$ and $n_7$. 
When we construct the representation for $u_4$ and \textbf{just use collaborative filtering relationship}, we could take  $n_5$'s neighbors $u_3$, $u_2$ and $u_1$  as the 2-hop neighbors of $u_4$ and find nothing from $n_7$ since there is not other neighbors except $u_4$. 
Then the representation of $u_4$ is built by aggregating the representations of 1-hop and 2-hop neighbors. 
As a result, the representation of $u_4$ obtains more influence from the popular news $n_5$ and the impact of $n_7$ is weakened.  

To overcome the limitations of representation-based and GNN-based approaches, we propose the content filtering enriched GNN framework for news recommendation (\ProName). In \ProName, both collaborative and content filtering information are captured: (1) We use traditional collaborative filtering approaches, such as GNN-based methods, to learn the representations based on the user-news interactions; (2) We propose \textbf{N}ews \textbf{E}xpanding (NE) module as Generator and \textbf{T}arget \textbf{A}ware \textbf{F}iltering  \& \textbf{A}ggregation (TAFA) module as Discriminator to learn the content filtering representations. 
Specifically, we utilize the NE to expand neighbors for news and their neighbors generated based on the high-order content-based similarity, and then we leverage the TAFA to identify the importance of the expanded neighbors in different content-views. 
The plain intuition is that users prefer the news with similar contents.

\begin{figure}[!t]
  \centering
  \includegraphics[width=0.4\textwidth]{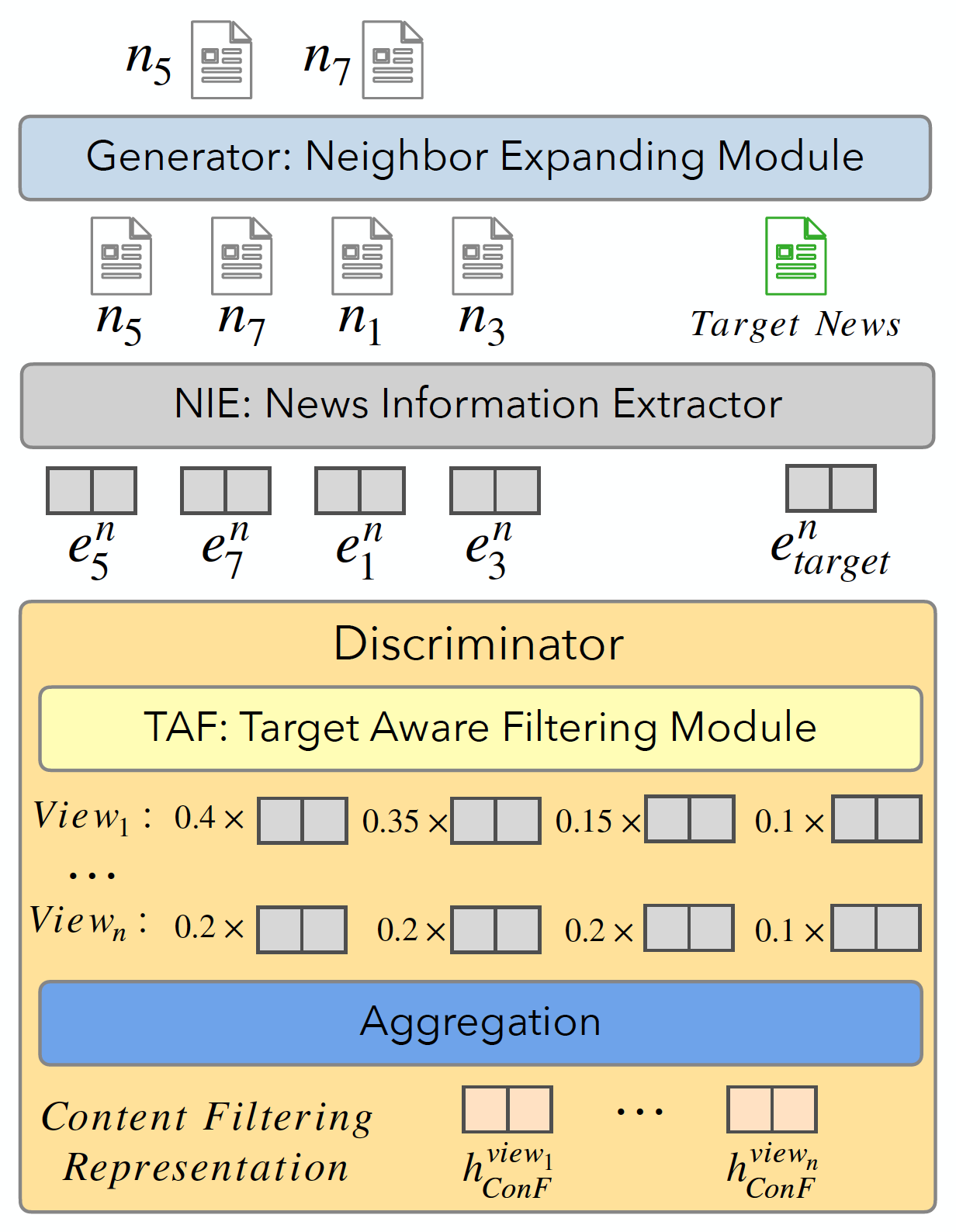}
  \caption{An example of content filtering representation component: Generator and Discriminator.}
  \label{fig:ConF_example}
\end{figure}

Following the example in Figure~\ref{CF_CF}, we give another example in Figure~\ref{fig:ConF_example} to introduce the process of building the content filtering representations for $u_4$. 
The user $u_4$'s behavior history includes $n_5$ and $n_7$, we can expand $u_4$'s neighbor as $n_5,n_7,n_1,n_3$ through Generator (NE module). 
After getting the embedding of news through NIE module, we discriminate the importance of these news under different content views through the TAFA module. 
Finally, the content collaborative representation for user $u_4$ is aggregated from the representations of news in the expanded list. 
As the procedure is similar, we omit the construction process of content filtering representations for news due to the space limit. 
Based on NE and TAFA modules, both user's and news' data sparsity problems can be alleviated.
The contributions are summarized as follows:
\begin{itemize}[leftmargin=*]
    \item In this paper, we propose \ProName~framework to learn the representations of user and news effectively. 
    The proposed framework improves the recommendation performance by fully considering both collaborative and content filtering information, and is compatible with existing GNN-based approaches for news recommendation.
    \item The proposed framework is able to alleviate the data sparsity problem. Specifically, the NE module is the generator to generate neighbors for news and the TAFA module is the discriminator to identify important news.
    \item We conduct comprehensive experiments and compare the state-of-the-art baseline models for news recommendation to demonstrate the effectiveness of the proposed method. Moreover, to clarify the contribution from different modules, the ablation study and case study are presented.
\end{itemize}

\section{Related Work}

In this section, we will review relevant research on news recommendation tasks.

With the explosion of a gigantic number news articles, to make better personalized news recommendation based on the user’s interests has been widely explored in recent years, and has wide applications~\cite{zheng2018drn,wu2019neural}. 
Early works~\cite{liu2010personalized, son2013location} used the manually designed features to obtain meaningful news and user representations. 
However, The above methods heavily rely on expert knowledge. 
To capture more informative knowledge with the end-to-end manner, the deep learning based representation approaches~\cite{wang2018dkn,NPA,NRMS} are proposed. 
Wang et al.~\cite{wang2018dkn} proposed DKN to leverage the knowledge modeled by the knowledge-graph. 
Wu et al.~\cite{NPA} proposed a news recommendation model with personalized attention mechanism to select important words and news articles based on user preferences to learn more informative news and user representations. 
Wu et al.~\cite{NRMS} proposed a neural news recommendation method which uses multi-head self-attention to learn news representations from the words in news and learn user representations from user's click history.

Recently, graph neural network (GNN) is widely used in recommendation~\cite{GAT, lightgcn,PinSage,BGCF}  because of its powerful representation ability for node features and graph structure. Compared to traditional matrix factorization based approaches, such as MF~\cite{koren2009MF}, GNN-based approach is able to capture high-order collaborative information.
Compared to traditional graph based approaches, such as label propagation~\cite{Bengio06LP}, GNN-based approach can capture non-linear features. 
Thus, several GNN-based approaches~\cite{GNUD,GERL} for news recommendation are proposed as the representation-based approaches.  
For instance, Hu et al.~\cite{GNUD} proposed GNUD which uses a GNN to capture high-order collaborative information. 

However, almost existing news recommendation methods, either representation-based or GNN-based, heavily rely on the collaborative filtering signal, namely the user's behavior history and user-news interactions respectively. But these methods may achieve sub-optimal performance when the user's behavior history is short and the interactions are sparse.

\section{Preliminary}\label{sec:pre}

Assuming the data set $D$ for training consists of $|D|$ user-news historical interactions $[u,n,y_{u,n}]$, where $u$ indicates the user id and the related user information, $n$ means the target news id and its features. 
And $y_{u,n}\in\{0,1\}$ is the associated label indicating user click behavior ($y_{u,n}=1$ means the user $u$ clicked the target news $n$, and $y_{u,n}=0$ otherwise).
To simplify the explanation, we note $y_{u,n}$ as $y$.
The task of news recommendation is to build a prediction model $\hat{y}=Model(u,n)$ to estimate the probability of a user $u$ clicking a specific news $n$.

According to GNUD~\cite{GNUD}, in this paper, we consider the title $T$ and profile $P$ (including entities $E$ and their corresponding entity types $C$ from the news content) as features for a  news article $n$. 
Each news title $T=\{t_1, t_2, ..., t_m\}$  is a sequence consists of m words.
And each news profile $P$ includes a sequence of entities $E=\{\omega_1, \omega_2, ..., \omega_p\}$ and a sequence of entity types ${C} = \{c_1, c_2, ..., c_p\}$, where p is the number of entities (entity types). 
We denote the embedding of title, entity, entity type as $\mathbf{T} = [\bm{t}_1, \bm{t}_2, ..., \bm{t}_m] \in \bm{R}^{m \times n_1}$, $ \mathbf{E} = [\bm{w}_1, \bm{w}_2, ..., \bm{w}_p] \in \bm{R}^{p \times n_2}$, and $\mathbf{C} = [\bm{c}_1, \bm{c}_2, ..., \bm{c}_p] \in \bm{R}^{p \times n_2}$, respectively. 
Following~\cite{DAN}, we define the profile embedding $\mathbf{P}=[[\bm{c}_1; w_1], [\bm{c}_2; w_2], ..., [\bm{c}_p; w_p]]^{T} \in \bm{R}^{p \times 2n_2}$. 
Moreover, we define the embedding of user id as $\bm{e}_u \in \bm{R}^{n_u \times n_3}$. 
Note that all above mentioned embeddings are randomly initialized. 



\begin{figure*}[ht]
  \centering
  \includegraphics[width=0.96\textwidth]{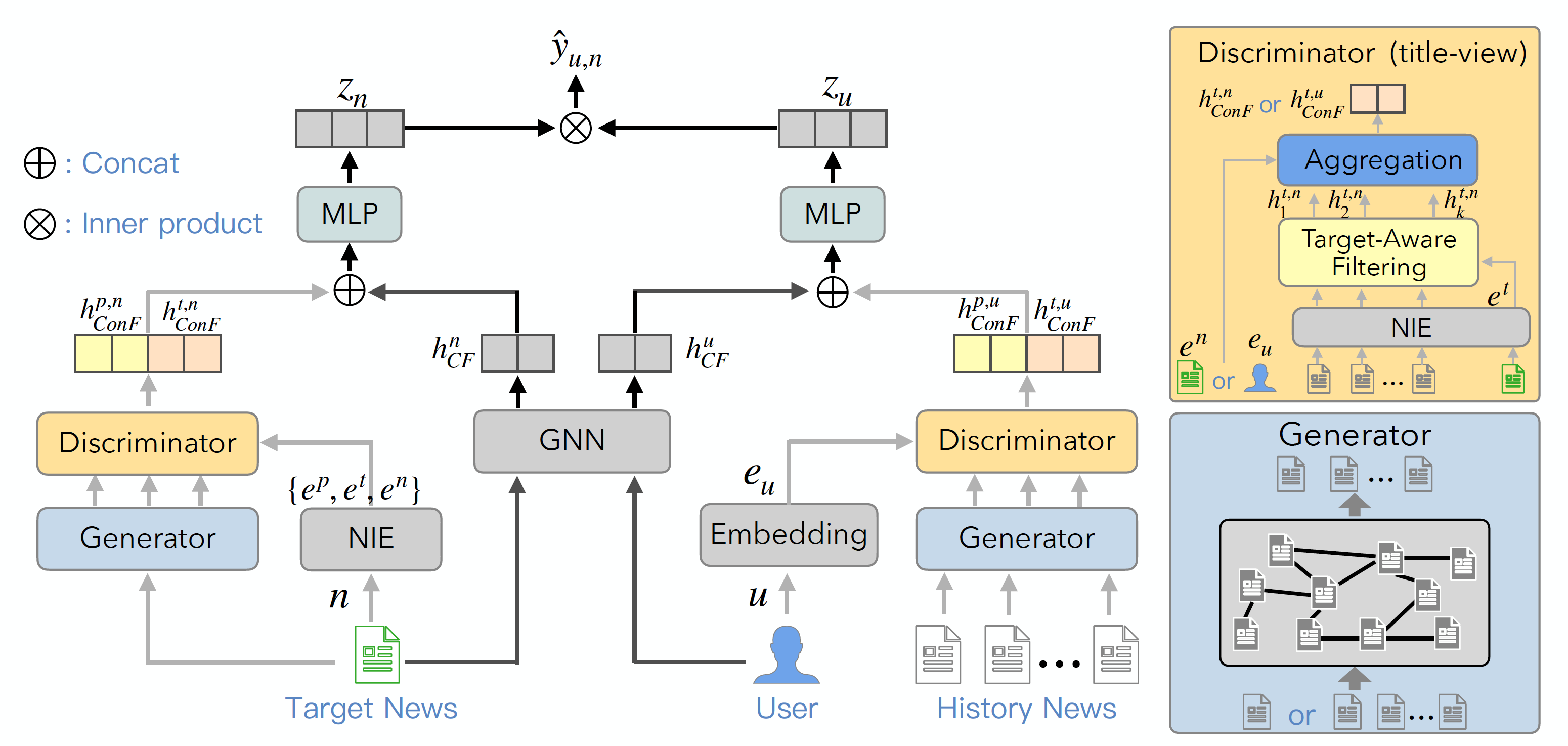}
  \caption{Illustration of the \ProName~framework. The "GNN" module means Graph-based neural network.
  }
  \label{GNEMF_framework}
\end{figure*}

\section{Methodology}
For the purpose of generating informative representations for user and news, we build user and news representations from collaborative filtering and content filtering.
As shown in Figure~\ref{GNEMF_framework}, the proposed \ProName~considers both collaborative and content filtering representations.
Specifically, the model is divided into two parts: the part one is based on traditional approaches (the GNN module in Figure~\ref{GNEMF_framework}), such as GAT and GNUD, and is used to capture the collaborative filtering information; the part two consists of
Generator and Discriminator, and is utilized to model content filtering information.
The Generator is a News Expanding (NE) module and the  Discriminator is Target Aware Filtering \& Aggregation (TAFA) module. 
In this section, we will give a brief description about the News Information Extractor (NIE) and collaborative filtering representation module since they are not the main contributions in this paper, and we will give a detailed introduction of \ProName.

\subsection{NIE: News Information Extractor}
News Information Extractor (NIE) is utilized to obtain a news representation $e^{n}$ from the raw news content, which consists of news title $T$, profile $P$. 
The raw content based representation would be taken as initial input embedding in the proposed framework.
Following~\cite{DAN,GNUD}, we also use two Parallel Convolutional Neural Network (PCNN) to encode news title ${T}$ and profile ${P}$ into the title-view and profile-view representations $\bm{e}^{t}, \bm{e}^{p}$ respectively. 
Finally, we concatenate $\bm{e}^{t}$ and $\bm{e}^{p}$, and get the original news representation $\bm{e}^{n}$ through a fully connected layer $f$:
\begin{equation}
\small
    \bm{e}^{n} = \bm{f}([\bm{e}^{t}; \bm{e}^{p}]).
\label{news_content_extractor}
\end{equation}

\subsection{Collaborative Filtering Representation}
Based on the original news representation $\bm{e}^{n}$ from raw content features, we can use graph based approaches, such as GNUD and GAT, to capture collaborative filtering information from user-news interactions and high-order relationships. 
Due to the space constraints, we do not elaborate on this module. 
The collaborative filtering representations of user and news are noted as $h^{u}_{CF}$ and $h^{n}_{CF}$.



\subsection{Content Filtering Representation}
As the example shown in Figure~\ref{fig:ConF_example}, the content filtering component consists of two major modules: 
(1) Generator:  leveraging content filtering to expand the user's history behavior and connect the news representations through the high-order similarity of news content;
(2) Discriminator: utilizing the features under different views of target news to identify the importance of news in user's behavior.
In this section, we will elaborate these modules in detail.

\subsubsection{Generator: News Expanding module}
In NE module, we first construct a news-news graph based on the content similarity between news, and then search top-k similar news from the news-news graph and finally obtain the k-nearest neighbor (kNN) graph  $\bm{G}_{kNN}=(\bm{A}_{kNN}, \bm{S}_{kNN})$, where $\bm{A}_{kNN}$ is the adjacency matrix, $\bm{S}_{kNN}$ is the similarity matrix, which is calculated by cosine similarity~\footnote{Other similarity metrics are also available and will be investigated in the future.}:
\begin{equation}
\small
    \bm{S}_{i,j} = \frac{ \bm{x}_i \cdot \bm{x}_j  }{ |\bm{x}_i| |\bm{x}_j| },
\label{similarity_computation}
\end{equation}
where $\bm{x}_i, \bm{x}_j$ are one-hot vectors of content for $i_{th}$ and $j_{th}$ news. Then, based on the similarity, we choose the top-k news pairs as the neighbors for each news and get the adjacency matrix $\bm{A}_{kNN}$.  

Further more, to explore the high-order similarity, based on $\bm{G}_{kNN}$, we follow~ \cite{perozzi2014deepwalk} and use Random-Walk to get the $\bm{G}_{walk}=(\bm{A}_{walk}, \bm{S}_{walk})$. 
For each news, taking itself as the starting node, we repeat the walk $ n_{walk} $ times, set the restart probability and the depth of each walker as  $p_{restart}$ and $d_{depth}$, respectively.
Then we get the Random-Walk similarity matrix between n items and select top-k similar node pairs for each news as its neighbors, and the adjacency matrix $\bm{A}_{walk}$ and similarity matrix $\bm{S}_{walk}$ are obtained consequently. 
Different from $\bm{G}_{kNN}$ which only relies on the local similarity, $\bm{G}_{walk}$ contains the global similarity which links more generalized news neighbors for each news. 

In our framework, based on $\bm{G}_{walk}$, we expand the target news and the user behavior sequence as $N^{n}_{+}$ and $N^{u}_{+}$, respectively.
For $N^{n}_{+}$, we just select the top-k similar news for target news. 
For $N^{u}_{+}$, we first search top-k similar items for each news of user clicked history, then de-duplicate all of them as a set, finally select the top-k similar news as the expanded neighbor set.


\subsubsection{Discriminator: Target Aware Filtering \& Aggregation module}
Equipping with NE module can expand user behavior sequence from content-view and boost the performance. 
However, utilizing NE module is likely to bring noise. 
Inspired by~\cite{ubr}, we propose the Target Aware Filtering (TAF) \& Aggregation module (TAFA in short) to identify importance of neighbor news.

\textbf{Target Aware Filtering:}
As introduced in Section~\ref{sec:pre}, there are two kinds of contents, title and profile. To identify the importance of different news under different views before aggregation, we adopt the multi-head attention mechanism to calculate the attention scores of the neighbors and the target news in different views.  For a news list $N$, we use the embedding $e^{t}$ and $e^{p}$ of target news $n$ to calculate the attention scores of the title and profile views, respectively.
For instance, the $j_{th}$ head's attention score of $i_{th}$ news of the neighbors on the target news $n$ in title view ($t$) is:

\begin{equation}
\small
    \bm{r}^{t}_{i,j} = \tanh([\bm{e}^{t}_{i} \bm{Q}^{t}_{j}; \bm{e}^{t}\bm{Q}^{t}_{j}] \bm{V}^{t}_{j}),
\label{soft_filter_score}
\end{equation}
\begin{equation}
\small
  \bm{\alpha}^{t}_{i,j} = \frac{exp(\bm{r}^{t}_{i,j} / \bm{\tau })}{\sum_{l\in N}{exp(\bm{r}^{t}_{l,j} / \bm{\tau})}},
\label{soft_filter_score_softmax}
\end{equation}
\begin{equation}
\small
   \bm{\alpha}^{t}_{i} = \sum_{j\in |head|} \bm{\alpha}^{t}_{i,j},
\label{soft_filter_score_sum}
\end{equation}
where the $\bm{Q}^{t}_{j}$ and $\bm{V}^{t}_{j}$ are the parameters in $j_{th}$ head of title view, $[;]$ means concatenation, $\bm{\tau}$ indicates the temperature parameter for softmax.
Then, based on the attention score, the $i_{th}$ news embedding $e^n_i$ is transformed by the title view attention score as:
\begin{equation}
  \bm{h}^{t}_{i} = \bm{\alpha}^{t}_{i} \bm{e}^{n}_i.  \label{filter_score_apply}
\end{equation}
The embedding of news $i$ is transformed by the profile view as $\bm{h}^{p}_i$. 
Therefore, for the expanded news set of target news $N^{n}_{+}$, we obtain two embedding lists in title and profile views as:
\begin{equation}
\small
\bm{H}^{t,n}=[\bm{h}^{t,n}_{1}, \bm{h}^{t,n}_{2}, ...,                      \bm{h}^{t,n}_{k}]
\label{filter_list_title}
\end{equation}

\begin{equation}
\small
\bm{H}^{p,n}=[\bm{h}^{p,n}_{1}, \bm{h}^{p,n}_{2}, ...,                      \bm{h}^{p,n}_{k}]
\label{filter_list_profile}
\end{equation}
For the expanded news set of user clicked history $N^{u}_{+}$, we obtain two lists $\bm{H}^{t,u}$ and $\bm{H}^{p,u}$.

\textbf{Aggregation:} We aggregate the transformed embedding list into the content filtering representations of user and news by the other two multi-head attention networks, respectively. 
Due to the space constraints, we only give the brief description of generating content filtering representations of news in title view:
\begin{equation}
\small
    \bm{h}^{t,n}_{ConF,j} = \sum_{i\in N^{n}_{+}}{\beta_{i,j}^{t}h_{i,j}^{t,n}},
\label{equ:agg_title}
\end{equation}
where $\beta_{i,j}^{t}$ is the attention importance of $i_{th}$ news' embedding in the list $H^t$ on the center node embedding $e^n$ in $j_{th}$ attention head. 
We can get the final aggregation representations in title view: $\bm{h}^{t,n}_{ConF} = [\bm{h}^{t,n}_{ConF,1}; \bm{h}^{t,n}_{ConF,2}; ...; \bm{h}^{t,n}_{ConF,head}]$.
Note that $\beta_{i,j}^{t}$ is calculated by the attention mechanism according to Equation~(\ref{soft_filter_score}), (\ref{soft_filter_score_softmax}) 
with different $Q$ and $V$. 
Similar to the title view, we can obtain content filtering representations of news in profile view as $\bm{h}^{p,n}_{ConF}$. 
For user node, we can get its  content filtering representations in title and profile views as $\bm{h}^{t,u}_{ConF}$ and $\bm{h}^{p,u}_{ConF}$, respectively.

\subsection{Final Layer}
Based on the graph-based approach and the proposed \ProName~module, we obtain the collaborative filtering representations $h^{u}_{CF}$ and $h^{n}_{CF}$ for user and news, and the content filtering representations $\bm{h}^{t,n}_{ConF}$ and $\bm{h}^{p,n}_{ConF}$, $\bm{h}^{t,u}_{ConF}$ and $\bm{h}^{p,u}_{ConF}$ for user and news, respectively. 
Then we concatenate these embeddings for user and news, and transform it to the final user and news representations by a multi-layer perception (MLP) respectively:
\begin{equation}    
\small
 \bm{z}_{u}= \mathcal{MLP}([h^{u}_{CF}; \bm{h}^{t,u}_{ConF};\bm{h}^{p,u}_{ConF}]),
 \label{item_representation}
\end{equation}
\begin{equation}    
\small
 \bm{z}_{n}= \mathcal{MLP}([h^{n}_{CF}; \bm{h}^{t,n}_{ConF};\bm{h}^{p,n}_{ConF}]).
 \label{item_representation}
\end{equation}

\subsection{Model Training}
Same as~\cite{GNUD}, we use the simple inner product to compute the click probability score, which is computed as: $\hat{y}_{u,n}= z^{T}_u \cdot z_n$. 
We define the following log-likelihood loss function for training sample (u, n) with the ground truth $y_{u,n}$ :
\begin{equation}
\small
    \bm{\mathcal{L}_1} =  -[y_{u,n}\ln{\hat{y}_{u,n}} + (1-y_{u,n}) \ln{(1-\hat{y}_{u,n}}) ]
\label{log_loss_func}
\end{equation}
where $\hat{y}_{u,n} = \sigma(s_{u,n})$.
Then we apply the l2 regularization to avoid overfitting and the overall training loss can be rewritten as:
\begin{equation}
\small
    \bm{\mathcal{L}} = (1-\lambda)\bm{\mathcal{L}_1} + \lambda||\Theta ||_{2},
\label{train_loss}
\end{equation}
where $\lambda$ is the regularization coefficient, $\Theta$ indicates the embedding parameters of user, item contents (title, profile) and PCNN parameters.

\section{Experiments}

\subsection{Datasets and Experimental Setting}
\subsubsection{Datasets}

Following DAN and GNUD, we conduct experiments on the real-world online news dataset Adressa~\cite{2017adressa_data}~\footnote{
http://reclab.idi.ntnu.no/dataset/} to evaluate the proposed framework.
We use three datasets named \textit{Adressa-1week}, \textit{Adressa-2week} and \textit{Adressa-10week}, which are extract 1 week, 2 weeks and 10 weeks logs in chronological order from the dataset, respectively.
Following GNUD \cite{GNUD,DAN}, we select user-id, news-id, timestamp, the title and profile of news to build our data sets.
We remove the stop words and filter out the words of low-frequency (less than 5). 
The statistics of  datasets are shown in Table \ref{tab:statistics_of_datasets}.  We also split all three datasets into three parts in chronological order and according to the ratio of 5:1:1:
(1) We use the first part to construct the user-news graph and users' clicked history; 
(2) The second part is used to build the training samples; 
(3) We randomly sample 20\% instances from the third part as validation set and regard the remaining as test set. 
Note that, we update the user history in training process same as DAN. 

\subsubsection{Experimental  settings}
To be fair, according to GNUD, we set the embedding size of user and news as 128, the batch size $B=128$, and use the random uniform distribution $U(-0.01, 0.01)$ to initialize the embedding.
And then we sample one item (that the corresponding user does not click) from the candidates set for each positive sample. 
In NE module, through the validation set, we set $k=30$, the restart probability $p_{restart}$ and repeated walk number $n_{walk}$ for each node as 0.19 and $10^5$, respectively.
In TAFA module, we set the number of heads and the output dimension per head as 4 and 32 for attention operator. 
We apply Adam~\cite{kingma2015adam} for model optimization. 
Then we use the validation dataset to tune the regularization coefficient $\lambda$ as $0.001$,  learning rate as $5\times 10^{-4}$ respectively. 
We adopt AUC and F1~\cite{GNUD} as the metric and use the F1 value as a reference for early-stop.


\begin{table}[t]
  \caption{Statistics of our datasets.}
  \label{tab:statistics_of_datasets}
  
\resizebox{\linewidth}{!}{
  \begin{tabular}{c|c|c|c}
    \hline
    \hline
    Number &  Adressa-1week & Adressa-2week & Adressa-10week    \\
    \hline
    \# user           & 615835   & 813634  &  2731135      \\
    \# news           & 20431    & 35380   &  75377        \\
    \# edges          & 2324905  & 3636685 &  17656916     \\
    \# Test smaples   & 505806   & 1154502 &  5794316      \\
    \# vocabulary     & 2965     & 4883    &  9755         \\
    \# entity         & 9243     & 15212   &  29226        \\
    \# groups         & 14       & 14      &  14           \\
    \# average words    & 6.22     & 6.20    &  6.26       \\
    \# average entities & 20.81    & 20.60   &  20.03      \\
  \bottomrule
\end{tabular}
}
\end{table}

\begin{table*}[h]\centering
  \caption{The performance of different methods on news recommendation.}
  \label{result_performence}
   \resizebox{1.0\linewidth}{!}{
  \begin{tabular}{l|l|cc|cc|cc}
    \hline
    & \multirow{2}{*}{Models}& \multicolumn{2}{c|}{Adressa-1week} &   
        \multicolumn{2}{c|}{Adressa-2week} &
        \multicolumn{2}{c}{Adressa-10week}\\
    \cline{3-8}
     &  &  AUC   & F1     &      AUC & F1     & AUC      & F1  \\
    \hline\hline
\multirow{4}{*}{Traditional}  
    &LR        & 0.5939 & 0.3668 &   0.5919 & 0.4288 &   0.5471 & 0.3011  \\
    &DSSM      & 0.7666 & 0.7378 &   0.7399 & 0.6900 &   0.6791 & 0.6781  \\
    &WideDeep  & 0.7796 & 0.6892 &   0.7401 & 0.6323 &   0.6441 & 0.5343  \\
    &DeepFM    & 0.7742 & 0.7310 &   0.7512 & 0.6869 &   0.7015 & 0.6680  \\
    \hline
\multirow{2}{*}{Representation-based }  
   & FIM       & 0.7508 & 0.7198 &   0.7232 & 0.6810 &   0.6211 & 0.6682  \\
   & DAN       & 0.7866 & 0.7695 &   0.7485 & 0.7167 &   0.7197 & 0.7059  \\
    \hline
\multirow{3}{*}{Graph-based }  
  & GAT       & 0.8580 & 0.8309 &   0.8226 & 0.7914 &   0.8951 & 0.8792  \\
  & GERL      & 0.7831 & 0.7458 &   0.8283 & 0.7733 &   0.8829 & 0.7455  \\
  & GNUD      & 0.8665 & 0.8291 &   0.8861 & 0.8506 &   0.9023 & 0.8747  \\
  \hline
  \multirow{3}{*}{\ProName-based } 
   &GAT\_\ProName  & 0.8960 (\textbf{+3.80\%}) & 0.8646 (\textbf{+3.40\%})
                   & 0.9067 (\textbf{+8.40\%}) & 0.8582 (\textbf{+6.70\%})
                   & 0.9252 (\textbf{+3.00\%}) & 0.9027 (\textbf{+2.40\%}) \\
   &GERL\_\ProName & 0.8560 (\textbf{+7.29\%}) & 0.8146 (\textbf{+6.88\%})
                   & 0.8824 (\textbf{+5.40\%}) & 0.8412 (\textbf{+6.80\%})
                   & 0.9325 (\textbf{+5.00\%}) & 0.9106 (\textbf{+16.50\%}) \\
   &GNUD\_\ProName & 0.8927 (\textbf{+2.60\%}) & 0.8681 (\textbf{+3.90\%})
                   & 0.8996 (\textbf{+1.40\%}) & 0.8700 (\textbf{+1.90\%})
                   & 0.9121 (\textbf{+0.90\%}) & 0.8987 (\textbf{+2.40\%}) \\
               
               
  \bottomrule
\end{tabular}
}
\end{table*}

\subsubsection{Baselines}
To evaluate the effectiveness of \ProName,  we compare the state-of-the-art methods from three categories for news recommendation: traditional recommendation (LR, DSSM, WideDeep, DeepFM), representation-based (FIM, DAN) and graph-based (GAT, GERL, GNUD). The brief descriptions are introduced as follows:
\begin{itemize}[leftmargin=*]
\item \textbf{LR}~\cite{ftrl}: a generalized linear model that takes user-id, user's clicked news, and candidate news content (title, entity, and groups) as input.

\item \textbf{DSSM}~\cite{DSSM}: a deep structured semantic model. 
We model the user clicked news as query, candidate news as documents.

\item \textbf{Wide\&Deep}~\cite{wide-deep}: a widely used deep learning framework, which combines a linear  model and a deep model, for recommendation in the industrial scenario.
We feed the same feature as LR for linear part and the user's clicked news, the profile and the title  for deep part.

\item \textbf{DeepFM}~\cite{deepfm}: a general deep recommendation model that combines the factorization machines and deep neural networks.
We use the same input features as Wide\&Deep.

\item \textbf{FIM}~\cite{FIM}:
a fine-grained interest matching method, which hierarchically constructs multilevel representations with dilated convolutions for user's behaviors and target news. 
We use the same input as DSSM.

\item \textbf{DAN}~\cite{DAN}:
an attention-based neural network for news recommendation which uses a dynamic attention mechanism to model user historical behavior sequences.



\item \textbf{GAT}~\cite{GAT}:
a general GNN method with multi-head attention aggregator, using the user-news graph for news recommandation. Specifically, the initial embeddings of user node and news node are constructed from user id and news contents, respectively.

\item \textbf{GERL}~\cite{GERL}:
a news recommendation method with high-order user-news relatedness, which uses the transformer to build news semantic representations. We use the profile embedding as the topic embedding.

\item \textbf{GNUD}~\cite{GNUD}:
a deep graph neural model which maps user and news to k kinds of spaces for restriction constraints, and strengthens attention learning through iteration.
We use the same inputs feature as GAT.
\end{itemize}

The experimental settings of compared  baseline models are consistent with those in the original papers.
To ensure fair comparison, we use the same dimension and initialization method to initialize word, entity, and groups embedding.
For each experiment, we repeated it more than 5 times independently and reported the average results.



\subsection{Overall Performance}

The experimental results for news recommendation of different models on 1week, 2week and 10week datasets are shown in Table~\ref{result_performence},
where we have the following observations:

\begin{itemize}[leftmargin=*]
    \item  The deep learning based models achieve  better performance, since the deep learning technique is able to capture more non-linear information. The observation is from the fact that LR performs worse than the other models. 
    \item Attention mechanism is able to improve the performance. We can observe that DAN achieves better performance than compared traditional methods except DeepFM in terms of F1 on Adressa-2week.
    However, the performance of FIM is worse than traditional methods,  since the dilated convolutions structure to too complicated for the news information extraction since the news contents may short and sparse.
    
    
    \item Except GERL on 1week dataset, the graph-based methods achieve better performance than both traditional and representation-based methods due to capturing the high-order relationship between user and news. As shown in Table~\ref{result_performence}, the performance in terms of AUC and F1 is improved with a large margin. 
    
    \item Considering both collaborative and content filtering relationships, the proposed \ProName~framework is able to enhance the performance of compared graph-based in terms of both AUC and F1.
    As shown in the Table~\ref{result_performence}, comparing with GERL, the GERL\_\ProName~achieves an improvement of 5.0\% to 7.29\% in terms of AUC and 6.8\% to 16.5\% in term of F1, respectively.
    There are two possible reasons: 
    (1) \ProName~alleviates the data sparsity problem through expanding news with similar contents by NE module. Specifically, low-degree news obtains more chance to be trained. As a result, the user and news representations are improved;
    (2) \ProName~is able to aggregate more reasonable and accurate news information with the target-aware filtering attention mechanism.
    
    

\end{itemize}

\subsection{Ablation Studies}
In this section, we present the several ablation studies on both Adressa-1week and Adressa-2week datasets to explore the effectiveness of different modules~\footnote{The vanilla model indicates GAT if there is no special emphasis.}.
Firstly, we verify the effectiveness of NE and TAFA modules. 
As Table~\ref{tab:ablation_analysis} shows, both NE and TAFA modules in the proposed framework are demonstrated to be effective.
The NE module can boost the performance with expanding information for user's history and strengthen the relationship of news with closer feature similarity.
The TAFA module can improve the performance because the importance of items in different views are recognized.
Combining both modules leads to further improvement, indicating both feature based information expanding and reasonable target-aware filtering are necessary for the news and user representations.  



\begin{table}[!t]
\centering
  \caption{The effectiveness of NE and TAF modules.}
  \label{tab:ablation_analysis}
  \resizebox{0.84\linewidth}{!}{
  \begin{tabular}{l|cc|cc}
      \hline
     &
    \multicolumn{2}{c|}{Adressa-1week} & 
    \multicolumn{2}{c}{Adressa-2week}  \\
  
    \cline{2-5}
     &  AUC & F1 & AUC & F1   \\
    \hline
    Vanilla          & 0.8580 & 0.8309 &   0.8226 & 0.7914   \\
    With NE          & 0.8854 & 0.8510 &   0.8831 & 0.8506  \\
    With NE\&TAFA     & 0.8960 & 0.8646 &   0.9067 & 0.8582  \\
  \bottomrule
\end{tabular}
}
\end{table}

\begin{table}[!t]
\centering
  \caption{The effectiveness of High-order News-News similarity.}
  \label{tab:expanding_comparsion}
  \resizebox{0.84\linewidth}{!}{
  \begin{tabular}{l|cc|cc}
    \hline
     &
    \multicolumn{2}{c|}{Adressa-1week} &  
    \multicolumn{2}{c}{Adressa-2week}  \\
    \cline{2-5}
                      &  AUC  & F1      &      AUC &     F1   \\
    \hline
    Vanilla           & 0.8580 & 0.8309 &   0.8226 & 0.7914  \\
    $G_{kNN}$\_NE     & 0.8734 & 0.8478 &   0.8571 & 0.8351  \\
    $G_{walk}$\_NE   & 0.8960 & 0.8646 &   0.9067 & 0.8582  \\
  \bottomrule
\end{tabular}
}

\end{table}

\subsection{Effectiveness of High-order News-News Content Similarity} 

To demonstrate the effectiveness of high-order news-news content similarity, we conduct the experiments to compare the performance of kNN similarity ($G_{kNN}$) and random walk similarity ($G_{walk}$), on Adressa-1week and Adressa-2week datasets, respectively. From Table~\ref{tab:expanding_comparsion}, we find the walk-based NE achieves better performance than kNN-based NE, which mainly due to random walk can capture the global similarity to link much more high-order news neighbors in feature space.

\begin{figure}[!t]
  \centering
  \includegraphics[width=1.0\linewidth]{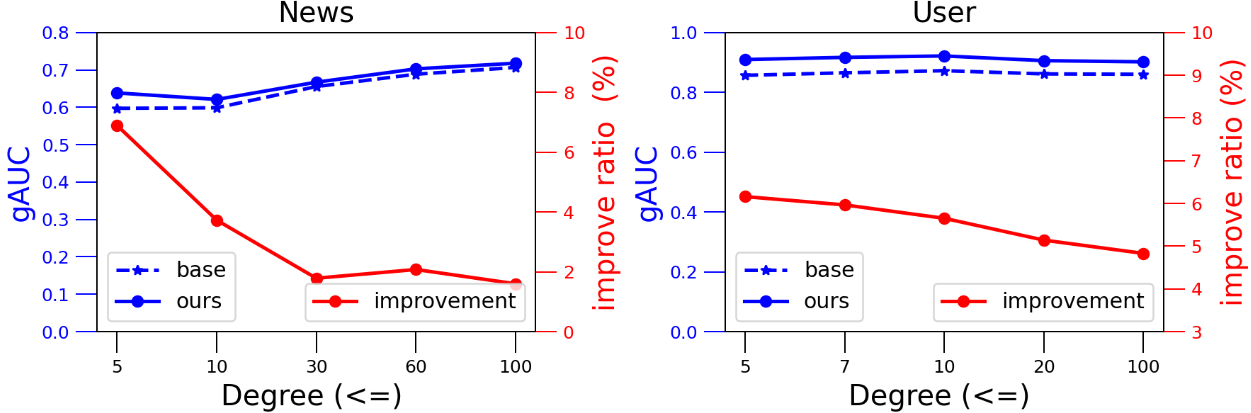}
  \caption{The effectiveness of \ProName~on data sparsity problem. The value on the horizontal axis represents the degree of node.
    The left vertical axis represents the gAUC and the right vertical axis indicates the relative improvement of \ProName~over vanilla. }
    
    
  \label{user_item_frequency}
\end{figure}

\subsection{Effectiveness on Data Sparsity Problem}

Figure~\ref{user_item_frequency} presents the effectiveness of the proposed \ProName~framework on the data sparsity problem. We take gAUC (group and average by user) on the Adressa-2week dataset as metric. 
In the left (right) of Figure~\ref{user_item_frequency}, each point indicates the result of the news (user) subset  where the degree (number of user/news neighbors) is smaller than the corresponding value in the horizontal axis. The results demonstrate the proposed framework can improve more for the users with short behavior sequence and the items with low popularity. In other words, the proposed \ProName~framework can effectively alleviate the data sparsity problem.

\begin{figure}[!t]
  \centering
  \includegraphics[width=0.98\linewidth]{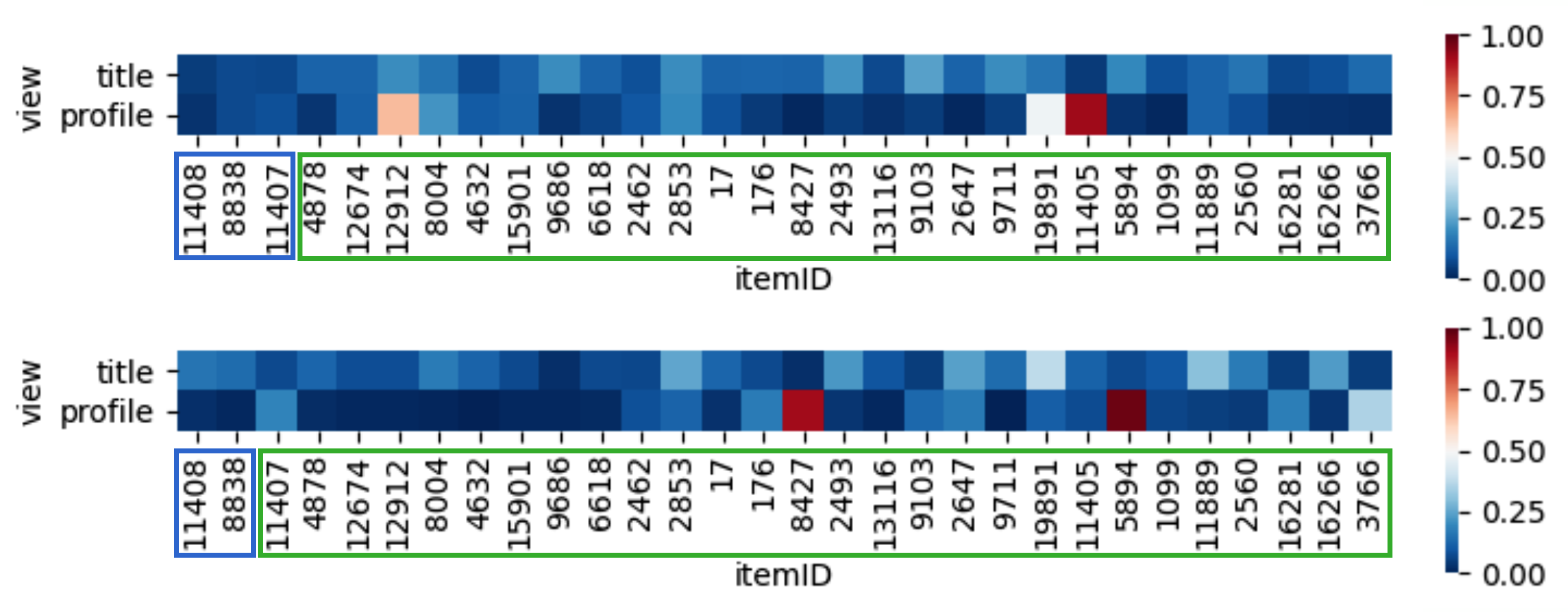}
  \caption{{The HeapMap of importance scores calculated by target-aware filtering. The horizontal axis represents news ID, vertical axis indicates two types of views, namely title and profile view, respectively. 
    }}
  \label{heatMap_filter}
\end{figure}
\subsection{Visualization of Target Aware Filtering}
For the purpose of elaborating the benefits of TAF, we random select two samples as example and visualize these distributions of filtering-based attention scores in user's Discriminator module.
As shown in the Figure~\ref{heatMap_filter}, each user's history $N^{+}$ consists of two parts: original clicked news (in the \textcolor{blue}{blue} box) and expanded news by NE (in the \textcolor{green}{green} box). 
The impact of embedding from different views are identified by TAF module.
The observations are from the fact that the different views of a same news have different importance.
Some of the expanded items have higher impact on target item which means that expanded information is more important.

\section{Conclusion}

In this paper, we propose the \ProName~framework for news recommendation. \ProName~is able to capture both collaborative and content filtering information and can alleviate the data sparsity problem. Specifically, we consider a feature-space similarity based News Expanding (NE) module to expand the news information for user history and target news as a generator, and utilize Target-Aware Filtering \& Aggregation (\TAFA~in short) module to identify the important clicked news and aggregate the item information like a discriminator. Comprehensive experiments are conducted on the real-world datasets. The results demonstrate the effectiveness of \ProName~to alleviate the data sparsity problem and improve the news recommendation performance.

\newpage
\bibliography{authology}
\bibliographystyle{acl_natbib}




\end{document}